# Considerations and recommendations from the ISMRM Diffusion Study Group for preclinical diffusion MRI:
# Part 3 — *Ex vivo* imaging: data processing, comparisons with microscopy, and tractography

Kurt G Schilling[1,2#], Amy FD Howard[3,4], Francesco Grussu[5,6], Andrada Ianus[7], Brian Hansen[8], Rachel L C Barrett[9,10], Manisha Aggarwal[11], Stijn Michielse[12], Fatima Nasrallah[13], Warda Syeda[14], Nian Wang[15,16], Jelle Veraart[17], Alard Roebroeck[18], Andrew F Bagdasarian[19,20], Cornelius Eichner[21], Farshid Sepehrband[22], Jan Zimmermann[23], Lucas Soustelle[24], Christien Bowman[25,26], Benjamin C Tendler[27], Andreea Hertanu[28], Ben Jeurissen[29,30], Marleen Verhoye[25,26], Lucio Frydman[31], Yohan van de Looij[32], David Hike[19,20], Jeff F Dunn[33,34,35], Karla Miller[4], Bennett A Landman[36], Noam Shemesh[7], Adam Anderson[37,2], Emilie McKinnon[38], Shawna Farquharson[39], Flavio Dell' Acqua[40], Carlo Pierpaoli[41], Ivana Drobnjak[42], Alexander Leemans[43], Kevin D Harkins[1,2,44], Maxime Descoteaux[45,46], Duan Xu[47], Hao Huang[48,49], Mathieu D Santin[50,51], Samuel C. Grant[19,20], Andre Obenaus[52,53], Gene S Kim[54], Dan Wu[55], Denis Le Bihan[56,57], Stephen J Blackband[58,59,60], Luisa Ciobanu[61], Els Fieremans[62], Ruiliang Bai[63,64], Trygve B Leergaard[65], Jiangyang Zhang[66], Tim B Dyrby[67,68], G Allan Johnson[69,70], Julien Cohen-Adad[71,72,73], Matthew D Budde[74,75], Ileana O Jelescu[28,76#]

[#]Corresponding authors — kurt.g.schilling.1@vumc.org, ileana.jelescu@chuv.ch

[1]Radiology and Radiological Sciences, Vanderbilt University Medical Center, Nashville, TN, USA, [2]Vanderbilt University Institute of Imaging Science, Vanderbilt University, Nashville, TN, [3]Department of Bioengineering, Imperial College London, London, UK, [4]FMRIB Centre, Wellcome Centre for Integrative Neuroimaging, Nuffield Department of Clinical Neurosciences, University of Oxford, Oxford, United Kingdom, [5]Radiomics Group, Vall d'Hebron Institute of Oncology, Vall d'Hebron Barcelona Hospital Campus, Barcelona, Spain, [6]Queen Square MS Centre, Queen Square Institute of Neurology, Faculty of Brain Sciences, University College London, London, UK, [7]Champalimaud Research, Champalimaud Foundation, Lisbon, Portugal, [8]Center of Functionally Integrative Neuroscience, Aarhus University, Aarhus, Denmark, [9]Department of Neuroimaging, Institute of Psychiatry, Psychology and Neuroscience, King's College London, London, UK, [10]NatBrainLab, Department of Forensics and Neurodevelopmental Sciences, Institute of Psychiatry, Psychology and Neuroscience, King's College London, London, UK, [11]Russell H. Morgan Department of Radiology and Radiological Science, Johns Hopkins University School of Medicine, Baltimore, MD, USA, [12]Department of Neurosurgery, School for Mental Health and Neuroscience (MHeNS), Maastricht University Medical Center, Maastricht, The Netherlands, [13]The Queensland Brain Institute, The University of Queensland, Queensland, Australia, [14]Melbourne Neuropsychiatry Centre, The University of Melbourne, Parkville, Victoria, Australia, [15]Department of Radiology and Imaging Sciences, Indiana University, IN, USA, [16]Stark Neurosciences Research Institute, Indiana University School of Medicine, IN, USA, [17]Center for Biomedical Imaging, NYU Grossman School of Medicine, New York, NY, USA, [18]Faculty of psychology and Neuroscience, Maastricht University, Maastricht, Netherlands, [19]Department of Chemical & Biomedical Engineering, FAMU-FSU College of Engineering, Florida State University, Tallahassee, FL, USA, [20]Center for Interdisciplinary Magnetic Resonance, National HIgh Magnetic Field Laboratory, Tallahassee, FL, USA, [21]Department of Neuropsychology, Max Planck Institute for Human Cognitive and Brain Sciences, Leipzig, Germany, [22]USC Stevens Neuroimaging and Informatics Institute, Keck School of Medicine of USC,

University of Southern California, Los Angeles, CA, USA, [23]Department of Neuroscience, Center for Magnetic Resonance Research, University of Minnesota, MN, USA, [24]Aix Marseille Univ, CNRS, CRMBM, Marseille, France, [25]Bio-Imaging Lab, Faculty of Pharmaceutical, Biomedical and Veterinary Sciences, University of Antwerp, Antwerp, Belgium, [26]µNEURO Research Centre of Excellence, University of Antwerp, Antwerp, Belgium, [27]Wellcome Centre for Integrative Neuroimaging, FMRIB, Nuffield Department of Clinical Neurosciences, University of Oxford, United Kingdom, [28]Department of Radiology, Lausanne University Hospital and University of Lausanne, Lausanne, Switzerland, [29]imec Vision Lab, Dept. of Physics, University of Antwerp, Belgium, [30]Lab for Equilibrium Investigations and Aerospace, Dept. of Physics, University of Antwerp, Belgium, [31]Department of Chemical and Biological Physics, Weizmann Institute of Science, Rehovot, Israel, [32]Division of Child Development & Growth, Department of Pediatrics, Gynaecology & Obstetrics, School of Medicine, Université de Genève, Genève, Switzerland, [33]Department of Radiology, Cumming School of Medicine, University of Calgary, Calgary, Alberta, Canada, [34]Hotchkiss Brain Institute, Cumming School of Medicine, University of Calgary, Calgary, Alberta, Canada, [35]Alberta Children's Hospital Research Institute, Cumming School of Medicine, University of Calgary, Calgary, Alberta, Canada, [36]Department of Electrical and Computer Engineering, Vanderbilt University,, [37]Department of Radiology and Radiological Sciences, Vanderbilt University Medical Center, Nashville, TN, USA, [38]Medical University of South Carolina, Charleston, SC, USA, [39]National Imaging Facility, The University of Queensland, Brisbane, Australia, [40]Department of Forensic and Neurodevelopmental Sciences, King's College London, London, UK, [41]Laboratory on Quantitative Medical imaging, NIBIB, National Institutes of Health, Bethesda, MD, USA, [42]Department of Computer Science, University College London, London, UK, [43]PROVIDI Lab, Image Sciences Institute, University Medical Center Utrecht, The Netherlands, [44]Biomedical Engineering, Vanderbilt University, Nashville, TN, [45]Sherbrooke Connectivity Imaing Lab (SCIL), Computer Science department, Université de Sherbrooke, [46]Imeka Solutions, [47]Department of Radiology and Biomedical Imaging, University of California San Francisco, CA, USA, [48]Department of Radiology, Perelman School of Medicine, University of Pennsylvania, Philadelphia, PA, USA, [49]Department of Radiology, Children's Hospital of Philadelphia, Philadelphia, PA, USA, [50]Centre for NeuroImaging Research (CENIR), Inserm U 1127, CNRS UMR 7225, Sorbonne Université, Paris, France, [51]Paris Brain Institute, Paris, France, [52]Department of Pediatrics, University of California Irvine, Irvine CA USA, [53]Preclinical and Translational Imaging Center, University of California Irvine, Irvine CA USA, [54]Department of Radiology, Weill Cornell Medical College, New York, NY, USA, [55]Key Laboratory for Biomedical Engineering of Ministry of Education, College of Biomedical Engineering & Instrument Science, Zhejiang University, Hangzhou, China, [56]CEA, DRF, JOLIOT, NeuroSpin, Gif-sur-Yvette, France, [57]Université Paris-Saclay, Gif-sur-Yvette, France, [58]Department of Neuroscience, University of Florida, Gainesville, FL, United States, [59]McKnight Brain Institute, University of Florida, Gainesville, FL, United States, [60]National High Magnetic Field Laboratory, Tallahassee, FL, United States, [61]NeuroSpin, UMR CEA/CNRS 9027, Paris-Saclay University, Gif-sur-Yvette, France, [62]Department of Radiology, New York University Grossman School of Medicine, New York, NY, USA, [63]Interdisciplinary Institute of Neuroscience and Technology, School of Medicine, Zhejiang University, Hangzhou, China, [64]Frontier Center of Brain Science and Brain-machine Integration, Zhejiang University, [65]Department of Molecular Biology, Institute of Basic Medical Sciences, University of Oslo, Norway, [66]Department of Radiology, New York University School of Medicine, NY, NY, USA, [67]Danish Research Centre for Magnetic Resonance, Centre for Functional and Diagnostic Imaging and Research, Copenhagen University Hospital Amager & Hvidovre, Hvidovre, Denmark, [68]Department of Applied Mathematics and Computer Science, Technical University of Denmark, Kongens Lyngby, Denmark, [69]Duke Center for In Vivo Microscopy, Department of Radiology, Duke University, Durham, North Carolina, [70]Department of Biomedical Engineering, Duke University, Durham, North Carolina, [71]NeuroPoly Lab, Institute of Biomedical Engineering, Polytechnique Montreal, Montreal, QC, Canada, [72]Functional Neuroimaging Unit, CRIUGM, University of Montreal, Montreal, QC, Canada, [73]Mila - Quebec AI Institute, Montreal, QC, Canada, [74]Department of Neurosurgery, Medical College of Wisconsin, Milwaukee, Wisconsin, [75]Clement J Zablocki VA Medical Center, Milwaukee, Wisconsin, [76]CIBM Center for Biomedical Imaging, Ecole Polytechnique Fédérale de Lausanne, Lausanne, Switzerland

# Abstract

Preclinical diffusion MRI (dMRI) has proven value in methods development and validation, characterizing the biological basis of diffusion phenomena, and comparative anatomy. While dMRI enables *in vivo* non-invasive characterization of tissue, *ex vivo* dMRI is increasingly being used to probe tissue microstructure and brain connectivity. *Ex vivo* dMRI has several experimental advantages that facilitate high spatial resolution and high signal-to-noise ratio (SNR) images, cutting-edge diffusion contrasts, and direct comparison with histological data as a methodological validation. However, there are a number of considerations that must be made when performing *ex vivo* experiments. The steps from tissue preparation, image acquisition and processing, and interpretation of results are complex, with many decisions that not only differ dramatically from *in vivo* imaging of small animals, but ultimately affect what questions can be answered using the data. This work concludes a 3-part series of recommendations and considerations for preclinical dMRI. Herein, we describe best practices for dMRI of *ex vivo* tissue, with a focus on image pre-processing, data processing and model fitting, and tractography. In each section, we attempt to provide guidelines and recommendations, but also highlight areas for which no guidelines exist (and why), and where future work should lie. We end by providing guidelines on code sharing and data sharing, and point towards open-source software and databases specific to small animal and ex vivo imaging.

**Keywords**: preclinical; diffusion MRI; ex vivo; best practices; microstructure; diffusion tensor; tractography; acquisition; processing; open science.

# 1 Introduction

The use of animal models and ex vivo tissue is essential to the field of diffusion MRI. In this work, **we define small animal imaging as imaging performed on a living experimental animal, whereas *ex vivo* we define as covering any fresh excised tissue, perfused living tissue, or fixed tissue.** Small-animal research is highly valuable for investigating the biology, etiology, progression, and treatment of disease; for the field of dMRI specifically, preclinical imaging is essential for methodological development and validation, characterizing the biological basis of diffusion phenomena, and comparative anatomy. While dMRI enables non-invasive characterization of tissue *in vivo*, *ex vivo* acquisitions are increasingly being used to probe tissue properties and brain connectivity. Diffusion MRI of *ex vivo* tissue has several experimental advantages, including longer scanning times and absence of motion. Together, these make it possible to acquire data with significantly higher signal-to-noise ratio (SNR), higher spatial resolution, and with sophisticated diffusion contrasts which may enable better characterization of tissue microstructure and structural connectivity. Another advantage afforded by *ex vivo* dMRI is the ability to compare diffusion data to histological data, bridging the gap between *in vivo* and histology for methodological validation. Because of these advantages, there have been an increasing number of dMRI studies on *ex vivo* tissue samples.

However, there are a number of considerations that must be made when performing *ex vivo* experiments.The steps from tissue preparation, image acquisition and processing, and interpretation of results are complex, with many decisions that not only differ dramatically from *in vivo* imaging of small animals, but ultimately affect what questions can be answered using the data. This work completes a 3-part series of recommendations and considerations for preclinical diffusion MRI. Part 1 focuses on small animal diffusion MRI[1], giving guidance for in vivo acquisition protocols and data processing. Part 2 presents best practices for dMRI acquisitions in ex vivo tissue covering hardware selection, fixation and sample preparation, MR scanning, and tissue storage. In this manuscript, Part 3, we discuss and give recommendations for everything that follows ex vivo acquisition: image pre-processing, diffusion quantification and/or model fitting, methodologies for comparisons with histology, and ex vivo fiber tractography. Finally, we give perspectives on the field, describing sharing of code and data, and goals that we wish to achieve. In each section, we attempt to provide recommendations and considerations, also highlight areas for which no recommendations exist (and why), and where future work should lie. An overarching goal herein is to enhance the rigor and reproducibility of ex vivo dMRI acquisitions and analyses, and thereby advance biomedical knowledge.

**This work does not serve as a "consensus" on any specific topic, but rather as a snapshot of "best practices" or "recommendations"** from the preclinical dMRI community as represented by the authors. We envision this work to be useful to imaging centers using small animal scanners for research, sites that may not have personnel with expert knowledge in diffusion, pharmaceutical or industry employees, or new trainees in the field of dMRI. The resources provided herein may act as a starting point when reading the literature, and understanding the decisions and processes for studying model systems with dMRI.

# 2 Data Pre-processing

In this paper we refer to *pre-processing* as steps that come before any diffusion fitting (tensors, biophysical models, etc.). Pre-processing thus includes data conversion (e.g. DICOM to NIfTI), noise reduction, artifact correction/mitigation or any step that aims at improving data quality. Processing refers to diffusion data fitting and normalization to standard space. See **Figure 1** for possible preprocessing steps.

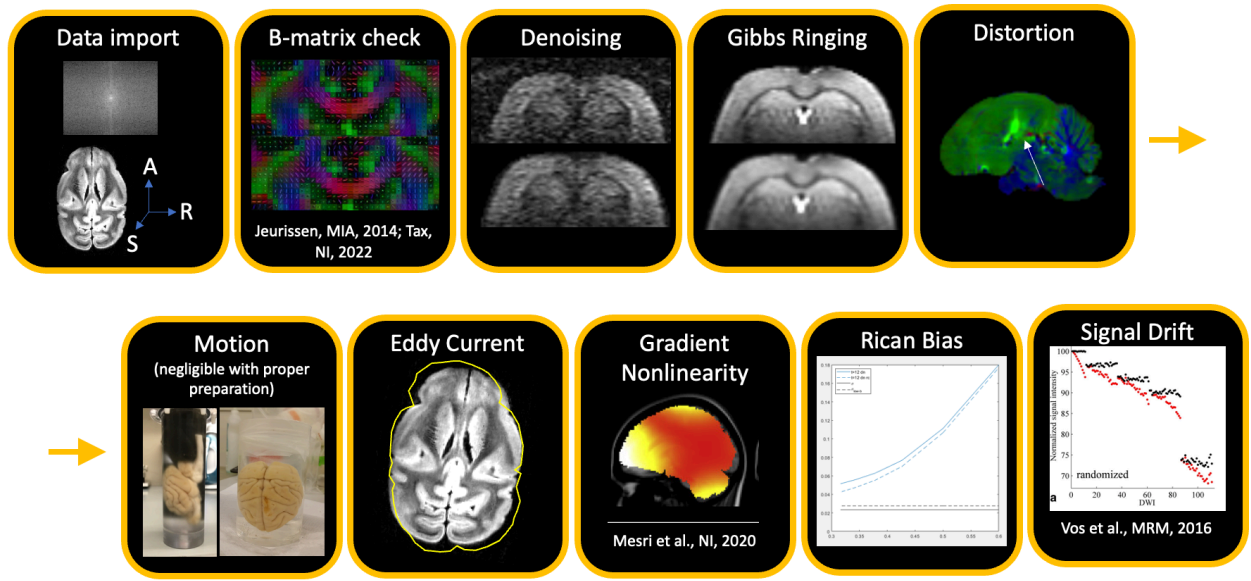

**Figure 1**. There are many artifacts that must be corrected for in preprocessing. These are not necessarily presented in order, and correction may not be necessary in all cases. Nevertheless, the most common order for pre-processing steps (after data important and quality check) is: (i) Thermal noise reduction (referred to as denoising), (ii) Gibbs ringing correction, (iii) Susceptibility distortion + motion + eddy current corrections (+ gradient non-linearity, if applicable), (iv) Rician bias correction and (v) Signal drift correction. Figures kindly provided by Ileana Jelescu, Kurt Schilling, or reproduced from [2,3].

Ex vivo diffusion MRI suffers from many of the same artifacts that in vivo imaging of both small animals and humans are susceptible to. This includes thermal noise, Gibbs ringing, signal drift, eddy current and susceptibility distortions, and sample motion. Below, we detail the steps associated with a typical pre-processing pipeline, stressing in particular what may differentiate implementations of *ex vivo* imaging from *in vivo*, and how available tools can/should be modified accordingly.

While most pipelines are designed for, and most heavily used in the brain, similar artifacts occur in all dMRI images. Our recommendations are generically applicable for all organs scanned ex vivo, although most details below are specific to brain imaging.

Any pipeline, regardless of sample, begins with **data importation and reconstruction**. While preclinical vendor software may output data in vendor-specific formats, common diffusion preprocessing software is most easily compatible with NIfTI or DICOM data, which stores not only the image matrix but also header information that includes information such as spatial resolution, sample orientation, and often acquisition parameters. With diffusion data, the diffusion weighting and diffusion directions are often stored as accompanying *b*-value and *b*-vector text files. The orientation of the images with respect to the applied diffusion directions is important, particularly for tractography, and should be quality checked carefully [4,5]. This is particularly important ex vivo where samples may be oriented differently to the living model. Consequently, header information which describes the orientation of the images (i.e. the Left-Right, Inferior-Superior, Anterior-Posterior labels) may need correcting and it is often common place to put some identifiable object (such as a fluid-filled capsule) or even a small physical cut in one hemisphere to ensure the hemispheres are correctly labeled. Tools for importation, reconstruction, conversion, and b-table quality control are given in [Section 5.1.1](#).

A prerequisite for many preprocessing steps is generating a **brain mask**. The mask is often used to save computation time or optimize areas on the image to focus corrections on. In general, brain mask generation can be challenging for small animal models because most digital "brain extraction" tools are optimized for and validated on the human brain. This can be more straightforward for ex vivo images, particularly if the sample was skull-extracted and immersed in perfluorocarbon fluids (e.g., fomblin) for imaging. One species agnostic approach for digitally extracting the brain from the image background that performs well on *ex vivo* images is to threshold the non-diffusion weighted image based on signal intensity, perform iterative dilation/erosion, and remove non-connected components[6]. When tissue is scanned within the skull, alternatively, species-specific strategies may be adopted[7,8]. One approach is to register all brains to a common space (template or atlas) where a brain mask is available or can be derived. The brain mask can then be registered back to native space, and optionally further adapted at the subject level[8].

Next, **denoising** aims to reduce thermal noise in diffusion-weighted images. Most denoising approaches and associated requirements are un-changed for *ex vivo* diffusion MRI. For example, a common approach based on principal component analysis (PCA) [9] and automated identification of signal and noise-carrying components (MP-PCA[10], NORDIC[11]) has proven useful in dMRI and is therefore recommended. The requirements here are that the noise level is constant across all diffusion images and that the number of diffusion images is large, where we suggest the use of >30 images. Other methods, for example total variation minimization [12] or non-local means denoising [12,13] are also applicable to *ex vivo* images. Several algorithms and packages are also provided in [Section 5.1.1](#). It is worth noting that most denoising methods assume Gaussian noise, whilst MR signal in magnitude images follows a Rician or non-Chi-squared distribution. Consequently, denoising should be ideally performed using complex or real-valued images to avoid enhancement of the rectified noise floor that may affect downstream modeling (see [14] for further discussion). We note that the MP-PCA algorithm can output a map of the noise level ($\sigma$) at each voxel, which can be used to calculate SNR maps for each diffusion-weighted image (by dividing the signal S by $\sigma$). A rule of thumb is that the SNR should be at least 2-5 for the highest b-value (note that it is also direction-dependent).

The next suggested step is **Gibbs ringing correction**, which is an artifact that appears as signal oscillations next to high contrast tissue interfaces and can interfere with model fitting for voxels near tissue edges e.g. in the corpus callosum[15], or cortex which are close to CSF. Gibbs ringing correction, while not dramatically affecting tractography, is important for microstructure modeling. Correction techniques include the methods described in [16] when a full Fourier acquisition is acquired and that of [17] when partial-Fourier acquisition is used (both for 2D multi-slice imaging). These methods have recently been extended to 3D[18], which is appropriate for the 3D acquisitions common in ex vivo imaging.

Next, **susceptibility distortions, eddy currents, and sample motion** need to be corrected. While susceptibility distortion in *ex vivo* scans can be mitigated through a multi-shot (segmented) acquisition, and motion should be minimal (with proper sample preparation), we still recommend correction for these potential artifacts. Pipelines and algorithms, such as those implemented within FSL (using the topup and eddy tools) or within TORTOISE (using the DR BUDDI tool) utilize a reverse phase encode scan to estimate the distortion field, and may use this field while correcting all three artifacts simultaneously. Regardless of software, care should be taken when using these pipelines with default parameters or configurations. For example, the knot-spacing or warp-field resolution for distortion/warping fields are typically set for human data acquired at ~2-2.5mm isotropic resolution, and should be scaled to match the possibly higher resolution ex vivo data under investigation. Additionally, practical compromises may need to be made (for example when choosing the number of iterations to run, or downsampling factors within the pipeline) for time considerations, particularly for ultra-high resolution datasets acquired with many diffusion-weightings.

**Rician bias correction** corrects the diffusion signal decay by subtracting the non-zero Rician floor that is present in magnitude data. Typical methods will assume the Gaussian noise standard deviation to be known, for example as previously estimated using MP-PCA on low *b*-value data. For software and methods, see [19,20] and Section 5.1.1. Alternatively, Rician noise models can be directly incorporated into the fitting procedure.

Finally, **temporal instability** on the scanner can cause signal drift, especially for diffusion sequences where strong gradients are employed for extended periods of time, even more so on preclinical scanners and especially for multi-day ex vivo experiments. This decrease in signal intensity over time can cause mis-estimates of derived parameters and also affect tractography [21]. Strategies to alleviate this effect include randomizing diffusion gradient directions and b-values, or better, explicitly designing direction sets to avoid consecutive directions with particularly heavy load on any one gradient axis. The presence of signal drift can be examined, and corrected by collecting multiple b=0 images throughout the scan to determine correction factors (typically linear or quadratic) to minimize this effect. Although this is not commonly done in the literature, we advocate for its use, and methodology and code to do so is described in [21] and in [Section 5.1.1](#).

# 3 Data Processing

Data processing includes fitting, normalization to a standard space, and tractography analysis. Diffusion analysis differences between *in vivo* and *ex vivo* are along the same lines as differences outlined for setting up the acquisition protocol (see Part 2), i.e., all changes are a direct result of potential differences/alterations in compartment sizes, diffusivities and relaxivities that are affected by chemical fixation and temperature.

## 3.1 DTI/DKI

To ensure that the assumptions underpinning DTI and DKI are valid, the b-values need to be set such that b×D ~ 1 (DTI limit), and b×D ~ 2-3 (DKI limit) [22] (where D is the diffusivity). This means that the maximum recommended *b*-values depend on the diffusivity and thus on the temperature at which the data were acquired. As described in Part 1[1], these guidelines result in a commonly used $b \simeq 1000$ s/mm$^2$ for DTI and highest b-value of $b \simeq 2000$-2500 s/mm$^2$ for DKI *in vivo*. As described in Part 2, ex vivo b-values should be increased by a factor of 2-5x depending on the drop in diffusivity. This means ex vivo DTI estimation can/should be performed based on b-values of $b \simeq 2000$-5000 s/mm$^2$, and the kurtosis tensor estimated from at least two shells with the highest b-value of $b \simeq 4000$-10,000s/mm$^2$.

There is clearly a wide range of possibly "optimal" b-values (we note that the lower end of these ranges are more typical in the literature). However, ex vivo also comes with the advantage that cursory scans should be used to investigate signal attenuation at different b-values for a given fixation and sample preparation procedure. The suitable b-value range for DTI and DKI analysis of a given ex vivo sample should be confirmed by examining the signal decay as a function of b-value to confirm the range of linear behavior (DTI regime: ln(S)~-bD) and measurable curvature for kurtosis quantification (ln(S)~-bD+⅙*(bD)$^2$K). It should be noted that DKI estimation is affected by the choice of *b*-values and post-processing (fitting procedures).

## 3.2 Biophysical modeling

For **biophysical models**, recommended *b*-values for optimal accuracy and precision of parameter estimation should also be adjusted *ex vivo*. Again, b-values should often be 2-5x the in vivo counterparts to account for the equivalent drop in water diffusivity. Similarly, diffusion times may need to be adjusted to account for this slower water diffusion ex vivo. See *Part 2* for further discussion of ex vivo acquisitions.

At the parameter estimation level, priors on diffusivities should be adapted to match *ex vivo* values, as well as potential admitted bounds on parameter values and algorithm initialization values. A typical example of this is an 'ex vivo flag' in the original implementation and source code of the NODDI model [23] which changed the assumed fixed diffusivity from 1.7E-6 mm$^2$/s to 0.6E-6 mm$^2$/s. This assumes the ex vivo diffusivity to be ~1/3 of its in vivo value. However, as noted elsewhere, ex vivo diffusivities are highly sample (fixation) and temperature dependent, making "universal" assumptions about ex vivo diffusivities often unreliable. Importantly, biophysical models may need to be adapted dramatically by the exclusion of compartments related to CSF, and inclusion of additional compartments, such as the "dot" compartment (trapped water with extremely low diffusion coefficient, see previous Section 2.2 *Ex vivo*: Translation and validation considerations") [24], for which *in vivo* evidence is limited to the cerebellum [25,26] and *ex vivo* more widespread to the cerebrum [27,28], spinal cord [29] and optic nerve [30]. As it can be difficult to know *a priori* whether an additional dot compartment is justified *ex vivo*, and because it can be challenging to disentangle it from the Rician noise floor, best practice can include fitting multiple models (with and without the dot) and determining model selection via e.g. estimate plausibility (parameters within biological ranges), precision, and the corrected Akaike or Bayesian information criterion. Example DTI, DKI, and biophysical model parameters maps for ex vivo mouse brains are shown in **Figure 2**.

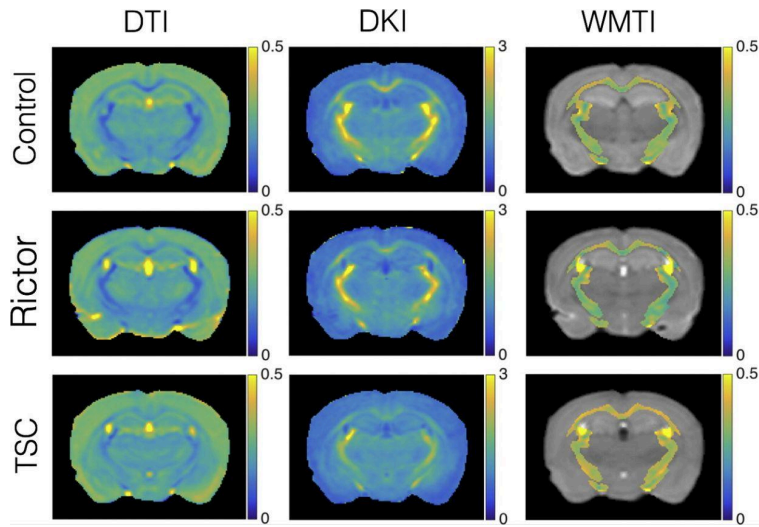

**Figure 2**. Ex vivo maps from DTI, DKI, and a biophysical model (white matter tract imaging; WMTI [31]. Parameter maps show radial diffusivity (DTI), radial kurtosis (DKI), and radial extra-axonal diffusivity (WMTI) for control mice and two hypomyelinated mouse models (Rictor and TSC). Ex vivo imaging was performed on a 15.2T Bruker Biospec scanner at 150μm isotropic resolution using a 3D diffusion-weighted fast spin-echo and b-values of 3000 and 6000 s/mm$^2$. Figure reproduced from [32]

## 3.3 Tractography

The application and use of fiber tractography as a tool to study the fiber pathways and wiring diagram of the brain remain largely the same for *ex vivo* (**Figure 3**) as for *in vivo* small animal and human dMRI, as fixation preserves the structure of axon bundles. In general, a measure of fiber orientation is estimated for each voxel, which is used to create continuous space curves (i.e., streamlines) which are thought of as representations of groups of axons traveling throughout the tissue. For these reasons, the fundamentals of tractography (deterministic and probabilistic algorithms) also remain the same, and guidelines follow that of human data.

For **acquisition**, we recommend acquiring data with isotropic resolution, as anisotropic voxel size can introduce bias in estimates of fractional anisotropy and hinder the ability of algorithms to deal with branching/bending pathways [33]. Higher angular resolution and strong diffusion weightings are likely to benefit tractography, particularly for small pathways, pathways near ventricles or gray matter boundaries, or pathways with high curvature. For most reconstruction techniques, we recommend acquiring greater than 30 diffusion-weighted directions (and commonly 60-100+, especially with little-to-no scan time limits). Example acquisitions with subsequent validation that have demonstrated reliable tractography results include the *ex vivo* mouse (0.1-mm resolution, 60 directions, $b$=5000 [34]), *ex vivo* ferret (0.24-mm resolution, 200 directions, $b$=4000 [35]), *ex vivo* squirrel monkey (0.3-mm resolution, 30-100 directions, $b$=1000-12'000 [35–37]), *ex vivo* macaque (0.25-mm resolution, $b$=4900, 114 directions [38]; 0.5-mm resolution, $b$=1477-8040, 180 directions [39,40]), and *ex vivo* pig (0.5-mm resolution, $b$=4000, 61 directions [41]) — all $b$-values given in units of s/mm$^2$.

The next step in the tractography process is estimating a **fiber orientation** for every voxel in the image. For ex vivo imaging, very little changes occur for this step, as most reconstruction techniques, including DTI [42], spherical deconvolution [43], ball & sticks models [44], and q-ball imaging [45], will result in a field of orientation estimates that can be used for tractography. As above, some fiber reconstruction methods may be adapted for ex vivo data through the inclusion of a dot-compartment to avoid estimation of spurious fiber orientations due to overfitting[46]. One important point to emphasize ex vivo is that several deconvolution methods may estimate a response kernel (the diffusion signal that results from a single fiber population) including an isotropic free water or CSF component [47], or may estimate a kernel for each tissue type - white matter, gray matter, CSF [48,49]. Because ex vivo tissue may not have CSF, or any free water if immersed in fomblin, care should be taken when using these algorithms to ensure they do not bias the true tissue components.

The **tractography** process itself is also largely unchanged ex vivo. As described in Part 1[1], it is still important to consider, and adapt, parameters that can be tuned. For example the step-size (the size of steps when propagating streamlines), curvature threshold (which stops streamlines if curvature is too high), or length thresholds (only allowing streamlines that are between a minimum and maximum total length). Adaptations should be considered based on acquired resolution, expected curvature of pathways under investigation, and length/size of the brain. For these reasons, most software packages for tractography (MRTrix3, DSI Studio, DIPY, FSL, ExploreDTI) are able to easily be used for *ex vivo* dMRI with few modifications.

Applications of tractography include bundle segmentation - the process of virtually selecting and dissecting pathways to study - and connectome analysis - assessing streamlines throughout the full brain to determine network properties, for example using graph-theoretic measures. Recommendations for these are identical to that for *in vivo* imaging (see Part 1[1]), where the primary challenges associated with small animals are the lack of automated bundle dissection tools in different species, and a lack of (or challenges in identifying) cortical parcellation schemes to use for connectome analysis.

Additional tractography applications involve the ability to study species beyond those conventionally used as scientific models. A few select examples include multiple primate brains for comparative anatomy and insight into brain evolution[50], studying auditory pathways in studies of dolphin brains[51], toxic exposure effects on connectivity (and parallels to temporal lobe epilepsy) in sea lions[52], or studying the extinct Tasmanian tiger brain[53] (preserved in formalin since 1905!) which have been extinct since 1936.

Finally, ex vivo imaging enables tractography in structures that may be challenging in vivo due to small size or motion. Examples include gray matter and intricate brainstem pathways in the human brain [54–58], or detailed mapping of the ascending/descending white matter, intra-cortical connections, and collateral fibers of the ex vivo spinal cord[59–61]. Outside the central nervous system, tractography has proven useful for characterizing normal and abnormal myofiber architecture and depicting sub-divisions of the ex vivo heart [62–64], or visualizing the course and structural abnormalities of ex vivo peripheral nerves[65–69], or tracing renal structures at high resolution in the ex vivo kidney[70]. While we do attempt to provide specific guidelines for these structures, we recommend strongly considering the goal of the tractography process in these locations (determining trajectory or orientation of tissue? Clustering structures? Measuring spatial extent of structures?) and how choices in the tractography process, including start/stop criteria, length and curvature thresholds, and streamline propagation methods may influence the ability to perform the desired tractography process.

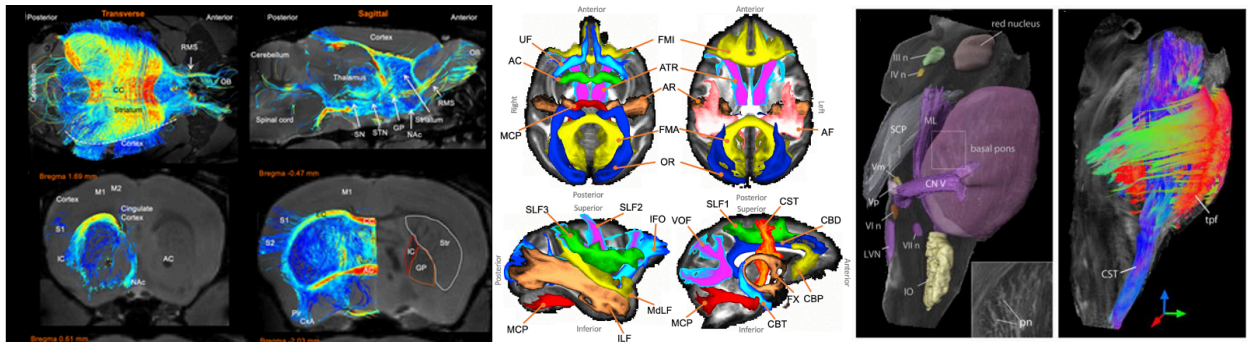

**Figure 3**. Ex vivo tractography on mouse (left)[71] , macaque (middle), and human brains (right). In the mouse, high resolution tractography was used to identify region-to-region differences in connectivity in models of Huntington's disease - here, tractography is able to delineate Striatal connectivity. In the Macaque brain, standardized protocols were developed to enable robust and automated segmentation of 42 white matter pathways[72]. In the human brain, diffusion data at high spatial resolution showed feasibility of reconstructing brainstem nuclei and white matter of the brainstem[73]

## 3.4 Normalization/Registration

It is common to use registration either to import atlas-based segmentation of brain regions (for ROI analysis or to use as tractography masks) or to bring individual maps into a common space for voxel-based comparisons. For this registration/normalization step, typical tools used in human data also work well for animal data, both in vivo and ex vivo, but often require some customization. For non-linear registration for instance, default physical dimensions of warp and smoothing kernels should be scaled to those of small-animal brains. Due to relaxation time and resolution differences in vivo versus postmortem, contrast between tissues may vary, and it can be challenging to non-linearly register ex vivo images to existing in vivo atlases (especially if the default cost-function is sum-squared-differences). In this case, it may be necessary to change the cost function (to mutual information, for example) or register to an explicit ex vivo atlas. Common MRI atlases, including brain segmentation, for a variety of species are provided in [Section 5.1.2](Section 5.1.2).

# 4 Comparisons with microscopy

## 4.1 Ex vivo MRI-microscopy comparisons

One of the main advantages of the ideal experimental conditions in *ex vivo* MRI (e.g., lack of motion and limited image distortions; high spatial resolution) is the possibility of deriving detailed microscopy information at accurate radiographic position [29,74–77]. This can then be used to validate MRI maps against microscopy indices quantifying similar biological features, or, more generally, to assess the correlation between MRI markers and a variety of microscopy-derived markers. An example of this is given in **Figure 4**, illustrating co-localised MRI and histological information from two published studies, i.e. i) in multiple sclerosis human spinal cord tissue [29] (top), and ii) in a mouse liver [78]. The figure shows how 2D microscopy from sections cut along a direction that is consistent with the MRI slice direction can be directly compared to MRI markers acquired in the same sample with good MRI-histology alignment [79], especially if 3D-printed molds customized to the specimen's anatomy are used to guide histological sectioning [80]. 3D microscopy is also possible [75,81], although it is usually limited to much smaller fields-of-view as compared to sample-wide 2D images, or requires very specialized protocols such as CLARITY [82].

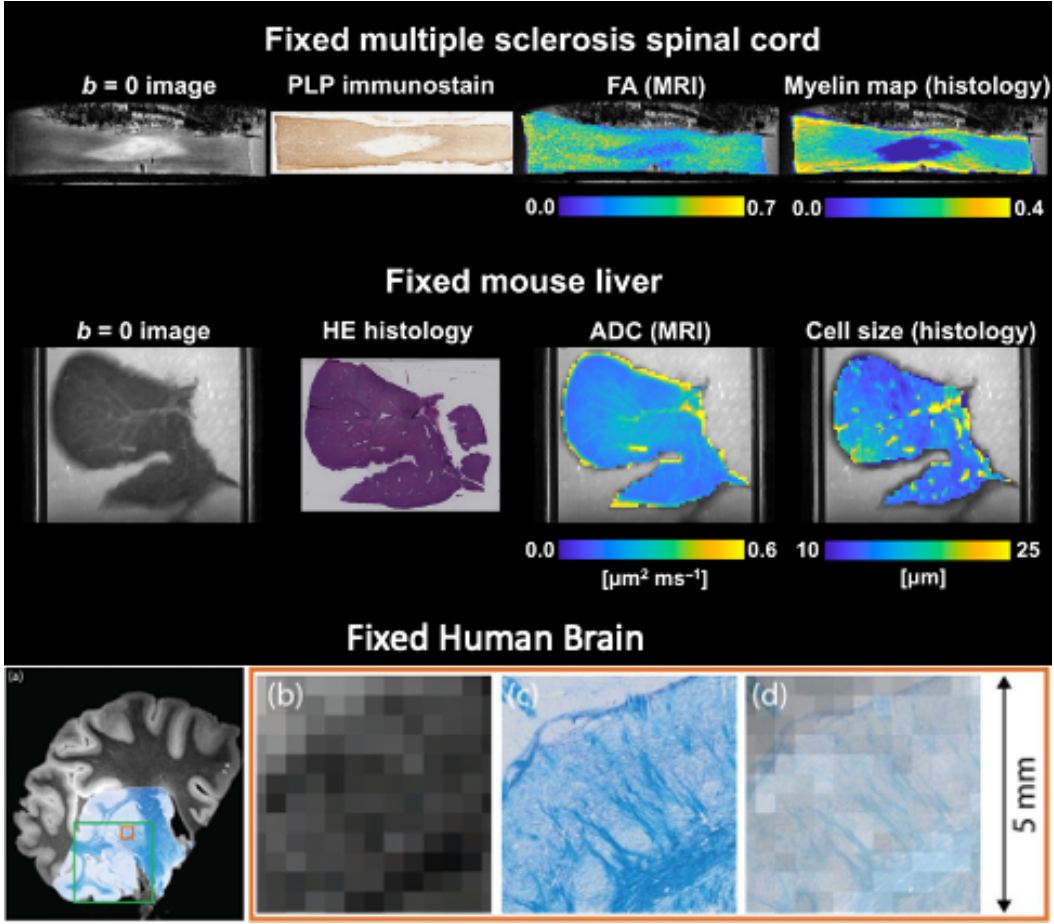

**Figure 4**. Examples of co-localised MRI and histological data. Top: fixed multiple sclerosis human spinal cord; bottom: fixed mouse liver. From left to right: b = 0 image; whole-sample histological section taken within the tissue corresponding to the MRI slice (proteolipid protein (PLP) immunostain for the spinal cord; hematoxylin and eosin (HE) staining for the mouse liver); dMRI parametric map (fractional anisotropy (FA) for the spinal cord; apparent diffusion coefficient (ADC) for the mouse liver); histological parametric maps co-registered to dMRI space (myelin staining fraction for the spinal cord, and volume-weighted cell size statistics for the mouse liver, evaluated within histological image patches matching the in-plane MRI resolution). The data reproduced in this figure with kind permission from C. A. M. Gandini Wheeler-Kingshott, G. C. DeLuca and R. Perez-Lopez refer to previous dMRI studies [29,78].

Irrespective of the chosen method, microscopy images are typically acquired at a resolution that is hundreds or thousands of times higher than the MRI voxel size. For example, typical resolutions on 2D slide scanner microscopes used for histology are of the order of 0.25-1.0 μm, while typical *ex vivo* dMRI resolutions are of the order of 100-500 μm. A common strategy to tackle the resolution mismatch is to combine the microscopy pixels that coincide with a given MR voxel into a "superpixel" or patch that matches the in-plane MRI resolution, and derive per-patch descriptors of microstructural properties (e.g., per-patch staining fractions, fiber orientation descriptors, or cell size distribution statistics [28,29,78,83]). This provides microscopy-derived parametric maps at a spatial scale that is comparable to that of diffusion MR images, to enable voxel-by-voxel MRI-microscopy comparisons.

For microscopy methods where the brain tissue is first sectioned into thin tissue samples, the resolution mismatch can not only relate to the in-plane resolution, but also to the slice direction: microtome cut thicknesses are of the order of 5-20 µm, while slice thickness in ex vivo dMRI is of the order of 200-1000 µm. This implies that a single microscopy slice only provides a partial picture of the microscopic characteristics underlying an MRI scan, since there are considerable portions of tissue that contribute to the observed MRI signals but that are not sampled. Better coverage can be achieved by imaging consecutive thin tissue slices that are then co-registered to create a 3D voxel volume[84], or by using 3D imaging methods with a relatively large field of view, such as advanced EM methods[85], confocal Imaging, 3D optical coherence tomography, or synchrotron-based tomographic imaging[86,87], or small-angle X-ray scattering tensor tomography.  However, some methods are only suitable for imaging smaller tissue samples, precluding whole-brain imaging of larger (e.g. primate) brains. Notably, very high-resolution imaging methods such as electron microscopy are often acquired in 3D from tissue blocks that are smaller than the dMRI voxel resolution, resulting in similar issues.

A further consideration is that while microstructure is inherently 3D (i.e cells are 3D objects and fibers have 3D orientations) some microscopy is much more informative in the 2D imaging plane (2D microscopy). These methods often provide a 2D projection of the structure of interest, or slice through it in the 2D plane, making inference on 3D microstructural metrics (such as fiber orientations or axon diameters[88]) difficult. This is an intrinsic limitation of studies relating 2D microscopy data to 3D MRI. One approach can be to take the diffusion metric (e.g., a 3D fiber orientation distribution) and similarly project it onto the 2D imaging plane (to create a 2D fiber orientation distribution), facilitating fair MRI-microscopy comparison[89–92].

In MRI-microscopy comparisons, sensitivity can be demonstrated using natural microstructural variation in healthy tissue, variation between pathological and control tissue, or via animal models in which specific tissue features can be purposefully manipulated (e.g. myelin manipulated through genetic modifications in shiverer mice or through environmental modifications in the cuprizone mouse model). In these cases, the correlation between ex vivo MRI and microscopy need not specifically match MRI voxels to microscopy data and a 1:1 correspondence between these contrasts may not be necessary.

## 4.2 Ex vivo MRI-microscopy alignment

There are several ways to **align MRI and microscopy** for quantitative comparison and validation. An excellent review of challenges and methodologies in registration of MRI to histology is provided in [93]. While not specific to ex vivo diffusion MRI, their recommendations and suggested protocols form the backbone of our review here, given in order of increasing technical difficulty.

First, the most simple, and arguably most common, approach is to manually select corresponding regions of interest in MRI and histology for quantitative analysis [94–96] especially suited for the small field-of-view of electron microscopy and similar techniques that are necessary for axon diameter and volume fraction quantification [24,32]. While no registration is required, it might be time consuming to manually select and delineate regions of interest in both microscopy and MRI, and a perfect correspondence is not guaranteed. For this reason, larger anatomical regions are typically selected from MRI (i.e., genu/body/splenium of corpus callosum, or large hand-drawn region of the cortex).

A second option is to aim to section the tissue specimen in planes parallel to the MR imaging planes. This will facilitate registration of 2D histology to the 2D MRI imaging plane using commonly employed registration packages, enabling a voxel-wise comparison of MRI and histology. A 3D printed mold may be created to facilitate registration. Here, an *in vivo* or *ex vivo* structural scan is quickly performed to create a 3D segmentation of the object (i.e. brain, prostate, spinal cord). Then, a mold is designed which not only holds the sample, but also has cutting guides, or slots, for cutting. Some guides may be nicely made to fit within specialized sample holders as well. Further scans can be performed *ex vivo*, where the FOV may be aligned with the cutting guide, so that there is a direct correspondence between the subsequent 2D histology and a slice (or slices) of the MRI image. This technique has been used for MRI imaging and histology alignment of multiple species and various organs [80,97–99], but, as of yet, not for diffusion validation directly.

A third option is to utilize an intermediate modality, usually referred to as block-face images, that are digital photographs of the tissue block as it is being sectioned. These block-face images can be registered individually, or first stacked into a 3D volume and registered to the 3D MRI. Because each 2D digital photograph can be mapped directly to a specific 2D histological slice, 2D registration can be performed to align histology to block-face (accounting for non-linear deformations arising from tissue processing), and subsequent 3D registration can be performed to align blockface to MRI (**Figure 5**). This technique has been performed in humans, mice, and monkeys, with dedicated pipelines and software [100–103], and has been shown to provide accurate alignment [104], and used to validate tractography, fiber orientation, and tissue microstructure measures. Registering 2D microscopy directly to 3D MRI is also possible in cases where block-face images were not acquired [105], though the optimization may be less well constrained.

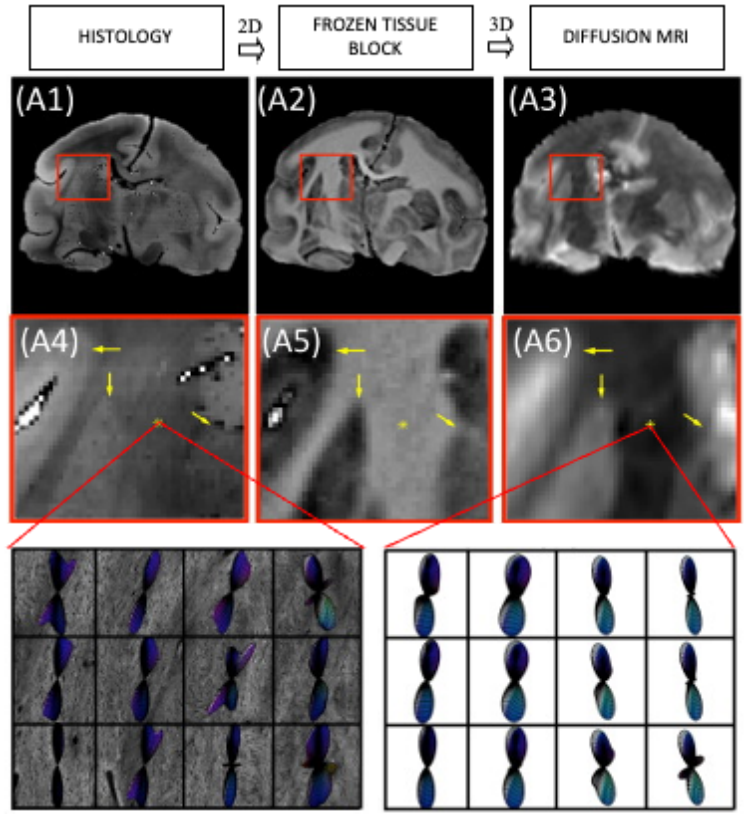

**Figure 5**. Histology to diffusion MRI alignment example using the intermediate modality, block-face images. 2D histology (A1) can be mapped to specific 2D block-face images (A2), which can be stacked into a 3D volume and mapped directly to 3D diffusion MRI data/derived data (A3). In this example, fiber orientation distribution from histology is aligned with similar measures estimated from diffusion MRI for validation purposes [76,81]. Insets (A4-A5) zoom in to show alignment across modalities, with final panels showing histology-derived (left) and MRI derived (right) fiber orientation distributions.

      We also note that in some special cases, especially with small samples imaged using planar surface coils, the entire RF coil, holder and sample can be removed for optical microscopy, providing a direct comparison as demonstrated on onion plant cells [106] and in mammalian brain slices (including human[96]), and muscle fibers [107–111]. Examples of this approach on a myelin-stained human spinal cord, Nissl-stained rat spinal cord, and rat hippocampal slice (with light microscopy) are shown in **Figure 6.**

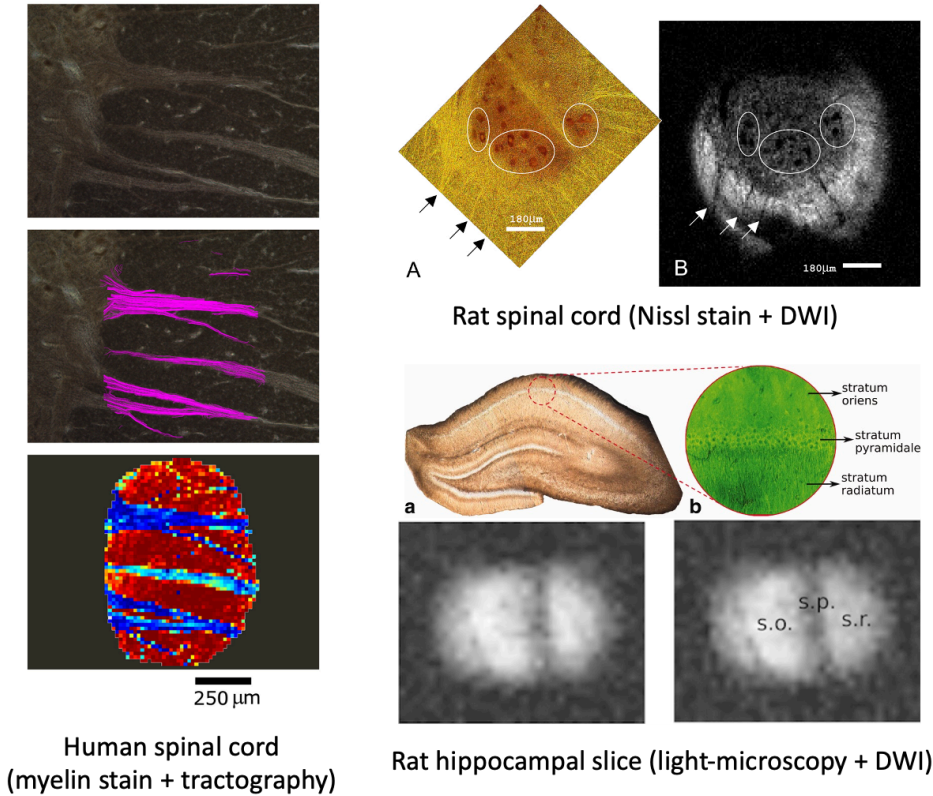

**Figure 6**. Examples of direct imaging of histological slices. (Left) A Black-Gold II stained histology (myelin-stain) was directly imaged in a microsurface coil with in-plane resolution of 15.6um. Tractography and red-green-blue orientation maps are shown overlaid on myelin stain. (Top Right) a 25um-thick Nissl-stained rat spinal cord tissue was imaged using a microsurface coil at 7.8um in-plane resolution with corresponding diffusion weighted image showing excellent correspondence. (Bottom Right) A 300um-thick rat hippocampal slice is imaged under a light-microscopoe and imaged using a slide-mounted microsurface coil at 12.5um in-plane resolution to study the microstructural effects of diffusion times and b-values. Images reproduced from [96] (left), [109] (top right), [111] (bottom right).

## 4.3 Optical imaging for validation

### 4.3.1 Histology

Histological staining is commonly used to visualize specific tissue features at µm-resolution. Ex vivo tissue is first sectioned into thin slices (typically ~5-10 µm), mounted onto glass slides and chemically stained. Many stains exist that target and visualize different tissue features. For example, there are common stains for myelin (e.g. PLP, Luxol fast blue or Gallyas/Bodian silver stains), Nissl or cell bodies (Cresyl violet, Golgi), neurofilaments (SMI-312), astrocytes (GFAP), microglia (Iba1), and many others. Some stains are chemical whilst others use antibodies to target specific proteins, known as immunohistochemistry. Stained slides are then imaged using optical microscopy with µm or sub-µm resolution where slide scanners are typically used for high-throughput 2D imaging. Digitized histology can be analyzed to extract microstructural metrics related to e.g. cell density, size, or the degree of axon myelination (via stain segmentation[112]) or fiber orientations (via structure tensor analysis[92,113]). The image processing involved needs to account for considerations that make stain (optical) density semi-quantitative: the stain density may not scale linearly with the antibody density and slides can suffer from artifactual staining variations both within and between slides. After processing, summary measures such as the cell count, the distribution of cell size, the number of stained pixels (stained area fraction) or the fiber orientation distribution can then be calculated over a local neighborhood and compared to dMRI-equivalents across regions of interest or on a voxel-wise basis. To separate sensitivity (showing a MR parameter correlates with some histology metric) from specificity (showing a MR parameter is selectively related to a single change in the tissue), multiple stains may be acquired and simultaneously analyzed to account for microstructural covariance across voxels/regions[112].

Histology (or immunohistochemistry) can also be combined with chemical tracers to enable precision mapping of axon trajectories from cortical regions of interest[114]. This form of neuroanatomical tract tracing provides "gold standard" estimates of brain connectivity that can be used to validate dMRI-based tractography (**Figure 7**). Tracer molecules are first surgically injected into a cortical region of interest where the tracer is taken up by neurons and actively transported along the axon, from the cell body to the axon terminals (anterograde tracers) or from the axon terminals to the cell body (retrograde tracers). Several weeks post-surgery, the animal is then sacrificed and the tissue sectioned and stained to visualize tracer deposition (i.e., stained cell bodies and axon trajectories) in the tissue. Sections sampled across the brain can then be digitized and combined, to map cortical-cortical or cortical-subcortical connectivity (i.e., injection/termination points), or create a 3D mask of axon projections across the brain. Tracers have been used extensively in animal models such as non-human primates and mice. However, tracers are typically limited to only one or two injection sites per animal, often requiring information to be combined across multiple animals, and tracers cannot be used in humans. Though it is possible to implant similar dyes in postmortem human samples[115], it can take a prohibitively long time for the dye to travel without active transport, meaning this method is only rarely used. Alternative tract-validation methods include gross white matter dissection, where fixed ex vivo tissue is first frozen and thawed (Kingler's technique) before being surgically dissected to reveal white matter fiber bundles[114], or comparisons of fiber orientations and downstream tractography from structure tensor outputs, or orientation-sensitive microscopy such as PLI, PS-OCT, SLI or SAXS, as described below and illustrated in **Figure 8**.

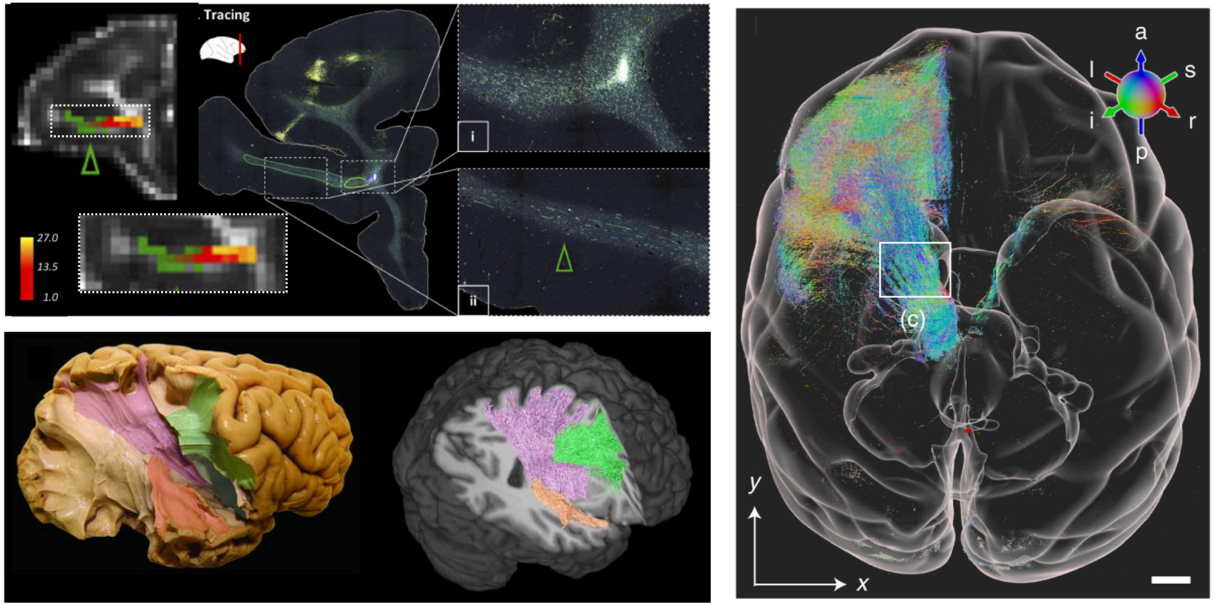

**Figure 7** Left: Tractography validation via anatomical tracers (top, [116]) and microdissection (bottom, [117]). Right: Recent advances in high throughput 3D imaging of anatomical tracers facilitates the tracking of single neurons, here projecting from the medial dorsal nucleus of the thalamus ([118]).

### 4.3.2 Fluorescence microscopy and tissue clearing

Fluorescence microscopy can similarly be used to identify specific tissue features with high resolution for single-cell-resolution imaging and analysis. Here, tissue is stained using fluorescent dyes and then imaged via more advanced optical methods such as confocal microscopy, light-sheet microscopy, 2-photon or super-resolution imaging. Multiple fluorophores can be labeled within the same tissue sample for co-localization of multiple tissue features, with the benefit that fluorescence can be directly related to fluorophore concentration given certain conditions, making the method often more quantitative than histological staining. Combining fluorescence with z-stack imaging, cellular morphologies and fiber orientations can be visualized in 3D for comparisons with dMRI[75,76,81].

Larger tissue sections (~mm³) can be imaged in 3D at depth by first optically clearing the tissue e.g. via CLARITY[119]. In tissue clearing, lipids are removed from the tissue such that the sample becomes optically transparent, facilitating 3D tissue imaging without sectioning. Penetration issues mean it can be challenging to both clear and label larger samples, though advancements in sample processing are ongoing[120]. Nonetheless, fiber orientations from cleared tissue have been successfully compared to diffusion MRI in both human and primate blocks of 10 × 10 × 0.5 mm³ [121], and the contributions to DTI of different cell types, myelin and fiber coherence disentangled in whole-brain mouse data[122,123].

### 4.3.3 Label-free imaging techniques

Label-free imaging methods utilize the intrinsic optical properties of tissue to generate contrast, without additional (exogenous) stains or dyes. Several techniques such as optical coherence tomography (OCT) and scattered light imaging (SLI) use the reflectance or scattering of light from tissue structures to drive contrast in the image. Analogous to ultrasound, OCT[124] uses an optical interferometer to obtain 3D depth-resolved images of reflected light at µm-resolution <~100 µm deep (depending on the sample). The top face of the sample is first imaged and then removed *in situ* (e.g. using a vibratome), and the process is repeated. This results in well-aligned images without the need for complex post-hoc registration. In comparison, scattered light does not image at depth, but illuminates the sample from different angles (typically from below, with light then transmitted through the sample) to estimate fiber orientations with an in-plane resolution of >6.5 µm [125,126]. The primary benefit of SLI over other methods lies in its ability to estimate multiple orientations per pixel in crossing fiber regions. Polarization sensitive methods such as polarized light imaging (PLI) or polarization-sensitive optical coherence tomography (PS-OCT) use the birefringent properties of the tissue to estimate orientational information (the optic axis) with micron-scale resolution [38]. As tissue birefringence in white matter is primarily related to myelin, fiber orientations can be inferred. The main difference between PLI and PS-OCT lies in the order in which the tissue is sectioned and imaged. In PLI[127], polarized light is transmitted through unstained tissue slices ~50-100 µm thick, whilst PS-OCT[128] uses reflected light and sections after imaging, as described in OCT above. As with SLI, setups often only provide reliable orientational information within the 2D imaging plane (the "in-plane angle"), though 3D PLI can be achieved e.g. through the use of a tilting sample stage[127]. SLI, PLI and PS-OCT have all been used to validate orientational information from diffusion MRI [56,91,105,125,129–131] (**Figure 8**).

### 4.3.4 Non-optical techniques

Non-optical techniques can provide benefits such as superior penetration or resolution to the optical methods above. High-resolution X-ray imaging (µ-CT or HiP-CT) can be used to image whole tissue samples at meso-scale resolution without sectioning, allowing 3D descriptions of cellular morphology and organization over considerable fields of view. This tomographic imaging method is analogous to clinical CT, but with high-energy X-rays providing µm-scale resolution ("µ-CT"): the 3D volume is constructed from 2D back-projections acquired as the tissue block is rotated through the x-ray beam[87]. Synchrotron x-ray sources can provide more intense and highly collimated x-rays facilitating super imaging compared to more typical "lab-based" µ-CT systems. To improve contrast, the tissue block is typically first stained with heavy metals such as osmium, though unstained phase-contrast methods available at synchrotron facilities (e.g. HiP-CT) are also gaining popularity[86,132]. In small angle X-ray scattering (SAXS), fiber orientations can be estimated from diffraction (Bragg) peaks in the X-ray scattering pattern due to myelin. SAXS provides a quantitative, myelin-specific signal with 3D fiber orientations and multiple, crossing fiber populations per pixel, though at more meso-scale resolutions of ~100 µm in-plane[125,133]. As x-ray imaging is non-destructive, µ-CT and SAXS can be combined with other contrasts such as electron microscopy or classical histology for multi-modal tissue investigations[87].

Electron microscopy can provide nano-scale visualizations of heavy metal (typically osmium) stained tissue to describe detailed cellular structures including cellular membranes, individual synapses, myelin lamella or features of the cytoskeleton. Tissue can either be imaged using back-scattered EM and then sectioned, preserving 3D localization of tissue structures, or first sectioned and then imaged and co-registered together (transmission EM), where the latter typically provides superior in-plane resolution[134]. EM samples are typically limited to small tissue blocks (~50x50x50 µm), though methods for high throughput, large FOV imaging are being developed[135].

As contrast in both µ-CT and EM is not cell-type specific, data analysis requires the post-hoc segmentation and identification of different cells or tissue (**Figure 8**). This can be challenging, though automated segmentation methods will continue to benefit from recent advances in machine/deep learning. EM-dMRI comparisons include validating variation in axon diameter across the brain, the degree of myelination for g-ratio mapping, or quantifying intracellular fractions. Further, 3D meshes from EM and µ-CT can be used as inputs for more microstructurally realistic simulations of water diffusion through tissue to investigate how deviations from highly simplistic tissue models favored in diffusion MRI (e.g. complex axon morphologies versus stick-like axons) impact the diffusion signal. Due to limited resolution and lack of contrast from relatively unmyelinated, low-diameter axons, µ-CT and EM can be biased towards large diameter axons, which may confound dMRI comparisons.

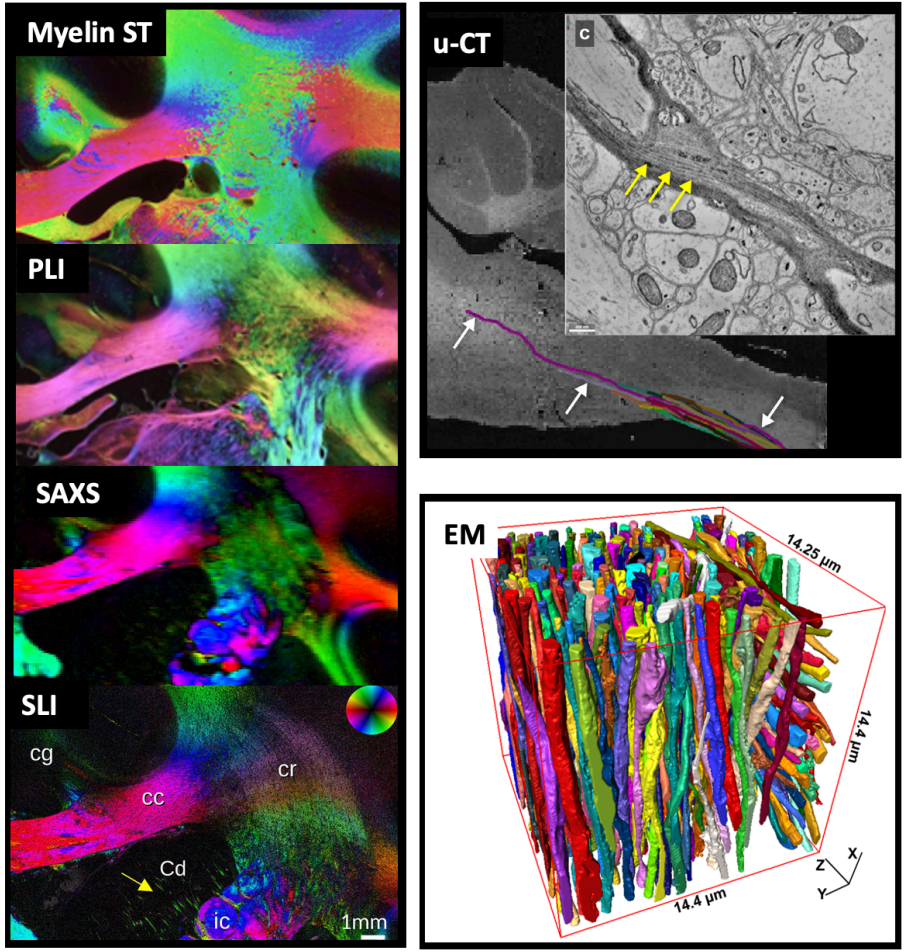

**Figure 8**. Fiber orientations extracted from various microscopy modalities can be used to validate biophysical models or fiber reconstruction methods in diffusion MRI. These modalities include structure tensor analysis of histological sections (myelin ST[105]) and polarized light imaging (PLI[105]), small angle x-ray scattering (SAXS[125]) and scattered light imaging (SLI[125]), micro-CT (u-CT[87]) and electron microscopy (EM[88]).

# 5 Perspectives

## 5.1 Open science

### 5.1.1 Code/Software

Challenges with pre-processing and processing pipelines highlighted in the previous sections could start to be overcome through code sharing and harmonization of implementations. Sharing combined knowledge and experience of many groups is valuable as it generates a lower barrier to entry and an excellent opportunity to evaluate robustness and reproducibility. We provide a (non-comprehensive) list of available software dedicated for acquisition and processing *ex vivo* diffusion MRI data at (https://github.com/Diffusion-MRI/awesome-preclinical-diffusion-mri) where updates on available software and tools can be shared by developers and where users can ask questions/advice for implementation, etc.

### 5.1.2 Data Sharing & Databases

Platforms that could serve as a repository for ex vivo dMRI datasets include OSF, OpenNeuro, Zenodo, NITRC, or other center resources (e.g. US National High Magnetic Field Laboratory). To promote data sharing and reuse, we compiled a (non-comprehensive) list of existing freely shared small-animal or *ex vivo* diffusion-weighted datasets, available on a public repository: https://github.com/Diffusion-MRI/awesome-preclinical-diffusion-mri.git. As for code sharing, the repository will enable a regular update of this database by the community.

## 5.2 The future: what should we strive to achieve?

As a field, we should continually strive to achieve reduced barriers to entry for new imaging centers, new scientists, and new industries who aim to use dMRI in a preclinical setting. Towards this end, as a community, we should promote dissemination of knowledge, code, and datasets to achieve high standards of data quality and analysis, reproducibility, transparency and foster collaborations, as well as reduce globally the time requirements and cost of research in this field. For easier and more direct translatability, direct access to and control over diffusion sequence timings on clinical systems would be a major benefit. A system for more organized sharing of custom sequences for preclinical systems would also lower the barrier for implementation of advanced techniques.

By design, *ex vivo* dMRI enables a more direct comparison / validation with invasive or destructive techniques such as histological stainings, chemical tracers, or electron microscopy. This potential can be exploited to its fullest to characterize and understand the biology behind a variety of diseases and injuries, thus contributing immensely to the translational value of dMRI.

Notwithstanding, for the translational circle to be complete, more research is needed to bridge *ex vivo* with *in vivo* measurements. So far, the extrapolation of *ex vivo* measurements to their *in vivo* counterpart has been hampered by open questions regarding the changes that the tissue undergoes during fixation and how those affect NMR-based measurements. Examples include the impact of partial volume effects between tissue types in high spatial resolution (*ex vivo*) imaging vs moderate resolution (*in vivo*) imaging, changes in compartment relaxation times and diffusivities, in membrane permeability, in relative compartment sizes, etc. Tissue fixation techniques such as cryofixation enable electron microscopy imaging of biological tissues where the *in vivo* structure was preserved to a greater extent that with regular chemical fixation [136] — which would provide a more realistic "ground truth" or comparative method for dMRI-derived *in vivo* microstructure, and for relative compartment sizes in particular. New preparation techniques have also recently enabled joint imaging using light microscopy (immunofluorescence) and electron microscopy on cryofixed tissue, with full hydration for light microscopy imaging [137]; the exploration of MR imaging of cryofixed rehydrated samples would certainly be worthwhile. Finally, with the advent of 3D large field of view microscopy with potential tissue clearing methods, it may be advantageous to perform direct 3D to 3D registration from histology to MRI for voxel-wise comparison and validation. However, due to large differences in resolution, contrasts, and geometric tissue distortions, substantial work is needed to make these comparisons feasible.

Future work related to preclinical imaging (and not specific to ex vivo) include:

**Pre-processing** steps are far from being optimized and integrated into a seamless pipeline for dMRI, so an initiative in this direction, ideally for each species, would highly benefit the community. We note this is not unique to ex vivo dMRI, nor preclinical dMRI, as there is no consensus or full understanding of the effects of different steps in preprocessing human *in vivo* data.

Transparent **processing** pipelines should also become the norm in the near future, though given the diversity and complexity of possible dMRI analyses, harmonization may be out of reach or even unjustified. We encourage new community members to search for existing tools in our GitHub database and expand/build on that.

New **biophysical models** of tissue are typically initially tested in a preclinical imaging setting. We underline that the development of new models should uphold high standards in terms of accuracy and precision of microstructural features estimated, and be validated using complementary techniques such as light or electron microscopy.

Rather than debate or controversy, most of the lack of **tractography** guidelines comes from a sparsity of resources dedicated to this application in the animal models. Future work could thus lie in creating resources that allow whole brain tractography in various species, followed by atlas-based labeling and bundle dissection for pathways of interest. As for biophysical models of microstructure, tractography is often validated in a preclinical setting. Thus, another path for future efforts is to understand and quantify differences between tractography and tracer, and to relate these to situations (i.e. tissue complexities such as crossing fibers) that may occur in the human brain.

To remain consistent with $b$-value units of $s/mm^2$ typically set at the scanner console and with "common language", we have reported b-values in $s/mm^2$ and diffusivities in $mm^2/s$ throughout this work. However, we would like to encourage the community to gradually adopt units that are more suitable for dMRI of biological tissue, where diffusion lengths are on the order of a few microns and diffusion times on the order of a few ms. Hence diffusivities expressed in $\mu m^2/ms$ and b-values expressed in $ms/\mu m^2$ are much more "natural" and enable to juggle numbers close to unity vs thousands (e.g. $b$=1 $ms/\mu m^2$ vs $b$=1000 $s/mm^2$) or decimals (e.g. $D$=1 $\mu m^2/ms$ vs $D$=$10^{-3}$ $mm^2/s$). Some of the recent literature on dMRI microstructure models have adopted this new convention, and we hope it will prevail in the near future.

# 6 Acknowledgements and Support

The authors acknowledge financial support from: the National Institutes of Health (K01EB032898, R01AG057991, R01NS125020, R01EB017230, R01EB019980, R01EB031954, R01CA160620, R01NS109090), the National Institute of Biomedical Imaging and Bioengineering (R01EB031765, R56EB031765), the National Institute on Drug Abuse (P30DA048742), the Secretary of Universities and Research (Government of Catalonia) Beatriu de Pinós postdoctoral fellowship (2020 BP 00117), "la Caixa" Foundation Junior Leader fellowship (LCF/BQ/PR22/11920010), the Research Foundation Flanders (FWO: 12M3119N), the Belgian Science Policy Prodex (Grant ISLRA 2009–1062), the µNEURO Research Center of Excellence of the University of Antwerp, the Institutional research chair in Neuroinformatics (Sherbrooke, Canada), the NSERC Discovery Grant, the European Research Council Consolidator grant (101044180), the Canada Research Chair in Quantitative Magnetic Resonance Imaging [950-230815], the Canadian Institute of Health Research [CIHR FDN-143263], the Canada Foundation for Innovation [32454, 34824], the Fonds de Recherche du Québec - Santé [322736], the Natural Sciences and Engineering Research Council of Canada [RGPIN-2019-07244], the Canada First Research Excellence Fund (IVADO and TransMedTech), the Courtois NeuroMod project, the Quebec BioImaging Network [5886, 35450], the Mila - Tech Transfer Funding Program and the Swiss National Science Foundation (Eccellenza Fellowship PCEFP2_194260), the Wellcome Trust (202788/Z/16/A, 203139/Z/16/Z and 203139/A/16/Z).
.